\documentclass[prc,showpacs,preprint,floatfix]{revtex4}
\include{psfig}

\begin{document}

\title
{\hfill{\small {\bf    }}\\
{\bf On the ill-posed character of the Lorentz integral
transform }\footnote[2]{Supported by the Deutsche Forschungsgemeinschaft (SFB
443). }
}
\author{W.\ Gl\"ockle}
\affiliation{Institut f\"ur theoretische Physik II,
 Ruhr Universit\"at Bochum, D-44780 Bochum, Germany}
\author{M. Schwamb}
\affiliation{Institut f\"ur Kernphysik, Johannes
Gutenberg-Universit\"at, D-55099 Mainz, Germany}

\begin{abstract}
\noindent
 An exact inversion formula for the Lorentz integral transform (LIT) 
is provided together with the spectrum of the LIT kernel. The
exponential increase of the inverse Fourier transform of the LIT
kernel entering the inversion formula explains the ill-posed
character of the LIT approach. Also the continuous spectrum of the
LIT kernel, which approaches zero points necessarily to the same
defect. A possible cure is discussed and numerically illustrated.


\end{abstract}

\maketitle

\section{Introduction}\label{kap1}
Integral transformations
\begin{equation}\label{int1}
L(y) = \int_{-\infty}^{\infty} d x \kappa(y,x) R(x)
\end{equation}
are widespread used in physics. According to Hadamard \cite{Had02}, the
problem of
 finding the unknown function
   $R$ for a given transform $L$ is called ``well-posed'',
 if the following conditions are fulfilled:
\begin{itemize}
\item The function  $R$ exists and is unique.
\item The  function  $R$ depends continously on  the input $L$.
\end{itemize}
Otherwise, the problem (\ref{int1}) is called ``ill-posed''. As example, let
 us consider the well-known  Laplace transformation
\begin{eqnarray}\label{laplace1}
L(y) = \int_0^{\infty} dx e^{-xy} R(x)
\end{eqnarray}
which is widely used in physics and engineering.
 It turns out to be ill-posed. For illustration, consider
 the  test function
\begin{equation}\label{laplace2}
R(x) = \frac{\sin(\frac{x}{\epsilon})}{\frac{x}{\epsilon}}
\end{equation}
and its Laplace transform
\begin{equation}\label{laplace3}
L(y) = \epsilon \arctan(\frac{1}{ \epsilon y}) \,\, .
\end{equation}
For $\epsilon \rightarrow 0$, the maximum
  norm of $L$ goes to 0, whereas the maximum
  norm of $R$ is 1 for any $ \epsilon$.

  In this paper  we study integral transforms, where the
  kernel $ \kappa $ in (\ref{int1}) has the form\footnote{Please 
 note that in the usual definition,
 no factor $\frac{\sigma_i}{\pi}$ occurs in  (\ref{LIT}).}

  \begin{eqnarray}\label{kernel}
  \kappa( y,x) = K(z=y-x)
  \end{eqnarray}
 In detail, we will concentrate on
  the Lorentz integral transform (LIT) 
($\sigma_i >0$ fixed)
\begin{eqnarray}\label{LIT}
K_{LIT} (z) = \frac{\sigma_i}{\pi} \frac{1}{  z^2 +
\sigma_i^2}
\end{eqnarray}
 which has recently been applied  extensively in photo- and electronuclear
 physics \cite{EfL07}.

  In the LIT-approach, due to the
choice $\sigma_i  \ne 0$, the calculation
 of physical observables in the $A$-particle scattering problem
 can be traced back rigorously to the solution of an
 appropriate bound state
 problem. This is of course a tremendous technical simplification
 and offers a unique possibility to carry out rigorous
 ab initio calculations on few-nucleon systems even
 beyond   mass number $A =4$
 \cite{EfL07}.
 However, the LIT is known to be ill-posed as will be again
 outlined below.

 In general,  in order to obtain
  $R$ from a given ill-posed  transform,  an appropriate
{\it regularization} scheme  is essential.
For example, in one class of inversion  schemes for the LIT, the
numerically gained
 Lorentz transform is expressed as a finite sum of appropriate basis functions
 whose inversion in explicitly known, i.e.\
\begin{equation}
L(y) = \sum_{i=1}^N c_i \chi_i(y) \,\, .
\end{equation}
Due to the ill-posed character of the LIT,   the upper value
$N$ of basis functions
 must be limited for obtaining reliable results. Otherwise, the gained
solution for $R$ may
 contain strong, unphysical oscillations. This reduction in the numerical
 resolution is nothing else as a regularization,
  see \cite{AnL05}
 for further details.

In this paper, we  intend to tackle the inversion problem of any ill-posed
 integral transform of type (\ref{kernel}) in a conceptually
 completely new manner, namely,  {\it without
 any use of regularization techniques.}
  For that purpose, we present
 in section 2  a new inversion formula for kernels of the
type (\ref{kernel}) which will then be applied to the LIT.  It turns out that
this inversion formula directly exhibits the
ill-posed character of the LIT. We further provide the spectrum of
that LIT kernel, which also makes its ill-posed character
evident.

Section 3 is devoted to a numerical case study for the LIT
approach, which illustrates the very source for the ill-posed
property. A possible reduction of that illness is proposed, i.e.\ a
 strategy will be developed to heal, at least under certain circumstances,
 the ill-posed character of the LIT.
  We end
with a brief summary in section 4.

\section{Mathematical properties of kernels including the LIT}

  For kernels of the type (\ref{kernel}),  eq.\ (\ref{int1}) has the form
  \begin{eqnarray}
  L(y) = \int_0^{\infty} dx K(y-x) R(x)
  \end{eqnarray}
  which Fourier transformed leads to
  \begin{eqnarray}\label{int3}
  \tilde L(k) = \sqrt{2 \pi} \tilde K(k) \tilde R(k)
  \end{eqnarray}
  with
  \begin{eqnarray}\label{Fourier_def}
\tilde f(k) = \frac{1}{ \sqrt{2 \pi}} \int_{-\infty}^{\infty} dy
e^{i ky} f(y)
\end{eqnarray}
 for any function $f$.
 Immediately, from (\ref{int3}), it
  follows the  inversion in {\it closed} form
\begin{eqnarray}\label{inversion}
R(x) = \frac{1}{2 \pi} \int_{-\infty}^{\infty} dk \frac{\tilde
L(k)}{ \tilde K(k)} e^{ - ikx}
\end{eqnarray}
Provided that the Fourier transform $ \tilde K(k)$ of the kernel
 is a continuous
function and that the infinum of its absolute value has a lower
bound larger zero, i.e.
\begin{equation}\label{int5}
{\cal C} = inf  |{\cal K}| > 0\,\, ,
\end{equation}
 we obtain for the $L_2$ norm
\begin{equation}\label{int6}
||R||^2 \equiv \int_{-\infty}^{\infty} dx |R(x)|^2 \leq \frac{1}{
4 \pi^2 C^2}
 ||L||^2\,\, . 
\end{equation}
In consequence,  we can formulate a very simple criterion to distinguish
 between ill-posed and well-posed integral kernels of the type
 (\ref{kernel}).
If ${\cal C} > 0$ is fulfilled,  the
 integral transformation is well-posed. Otherwise, it turns out to be
 ill-posed, as it becomes obvious by studying the spectrum  of integral
kernels of the type (\ref{kernel}). It is defined as
\begin{eqnarray}
\int_{-\infty}^{ \infty} dx K(y-x) \chi_{k_0} (x) =
\mu(k_0)\chi_{k_0} (y) \,\, ,
\end{eqnarray}
or Fourier transformed as
\begin{eqnarray}
\sqrt{2 \pi} \tilde K(k) \tilde \chi_{k_0} (k) =\mu(k_0)\tilde
\chi_{k_0} (k) \,\, .
\end{eqnarray}
This leads to
\begin{eqnarray}
 \tilde \chi_{k_0} (k) \sim \delta( k-k_0)
 \end{eqnarray}
 and
 \begin{eqnarray}
\mu(k_0) =\sqrt{2 \pi} \tilde K(k_0) \,\, .
 \end{eqnarray}

Therefore, the spectrum of any arbitrary integral kernel of the
 type (\ref{kernel}) is given by its Fourier transform. If a kernel fulfills
 ${\cal C} = 0$
 in (\ref{int5}), the  spectrum  is continuous  and has zero as
   accumulation point. In consequence, the kernel is  obviously
 ill-posed  and an   unprotected inversion cannot work.

  In case of the LIT,  $\tilde K(k)$ is given by
\begin{equation}\label{lor2}
{\tilde K}_{LIT}(k)= \sqrt{\frac{1}{2\pi}}  e^{-\sigma_i |k|} \,\, .
\end{equation}
 It  fulfills ${\cal C} = 0$,  its spectrum is continuous,
\begin{eqnarray}
\mu_{LIT}(k_0) = e^{ - \sigma_i | k_0|}
 \, \, ,
\end{eqnarray}
and approaches zero for large $k_0$. Therefore, the LIT is
 ill-posed.  Its  eigen functions are given by
\begin{eqnarray}
 \chi_{k_0} (x) \sim e^{ i k_0 x}
 \,\, .
\end{eqnarray}

\section{ A numerical case study for the LIT}

The application of the inversion formula (\ref{inversion}) requires a
 reliable  numerical treatment of Fourier transforms. In this work,
 this is performed  with the help of Filon's integration formula
\cite{AbS64}. In order to check our numerical routines, we take
a simple
 analytical test case for the function $R$:
\begin{equation}\label{lor5}
 R(x) = \sqrt{x} e^{-ax}  \Theta(x) \,\, ,
\end{equation}
 with $\Theta$ denoting the Heavyside  step function, i.e.\ the threshold
 of this function is placed at $x=0$. $a$ is a free parameter which we choose
 from now on as $a=0.05$ MeV$^{-1}$.  Following the nomenclature in literature,
 we call $R$ from now the response function.
Its Fourier transform
 is known analytically
\begin{equation}\label{fourier_ana1}
{\tilde R}(k)=  \sqrt{\frac{1}{8}}
 \, \left( \frac{1}{a^2+k^2}\right)^{\frac{3}{4}}
e^{i\frac{3}{2} \mbox{atan} \left(\frac{k}{a} \right)}
\end{equation}

 so that the quality of our numerical routines for the Fourier transform
 can be tested straightforwardly.
   For that purpose, we cut at first
  the integrations bounds
 in (\ref{Fourier_def})
 according to
\begin{eqnarray}\label{Fourier_def2}
\tilde R(k) = \frac{1}{ \sqrt{2 \pi}} \int_{-\infty}^{\infty} dx
e^{i kx} R(x) \longrightarrow \frac{1}{\sqrt{2\pi}}
\int_{-x_{max}}^{x_{max}}
 dx e^{i kx} R(x)
\end{eqnarray}

 with  $x_{max}= 3000$ MeV.  With respect to the strong exponential decrease
 of our test function  (\ref{lor5}), this upper value for $x_{max}$ is more
 than sufficient.  The remaining integral in
 (\ref{Fourier_def2}) is now evaluated with the help of Filon's integration
formula using $N$ equidistant grid points. It turns out that the numerical
 precision in calculating the Fourier transform increases naturally with
 increasing $N$,
 but decreases strongly with increasing argument $k$ of the Fourier
transform. For example, for $N=20001$ mesh points  the numerical error
 in the real part of the Fourier transform 
 is of the order of about 1.7  percent for  
$k = 0.8\, \mbox{MeV}^{-1}$, increasing to about 
 6.9  percent for  
$k = 2.0\, \mbox{MeV}^{-1}$.

At next, let us check  the inversion formula  (\ref{inversion}) for our
 test response (\ref{lor5}).  For that
 purpose, we  calculate at first numerically  the LIT transform 
$ L(y)$ of $ R(x)$ and then its  Fourier transformations
   ${\tilde L}(k)$ corresponding
   to (\ref{Fourier_def2}) with $x_{max}=3000$ and
 $N=20001$ equidistant mesh points. Then, with the help of the known
 Fourier transform of the Lorentz kernel (\ref{lor2}),
 we can perform the inversion (\ref{inversion}) where we again, similar as
 in (\ref{Fourier_def2}), have to cut the integration at the lower and upper
limit, i.e.\

\begin{eqnarray}\label{inverse2}
R(x) = \frac{1}{2 \pi} \int_{-\infty}^{\infty} dk \frac{\tilde
L(k)}{ \tilde K_{LIT}(k)} e^{ - ikx}
\longrightarrow
\frac{1}{2 \pi} \int_{-k_{max}}^{k_{max}} dk
\frac{\tilde
 L(k)}{ \tilde K_{LIT}(k)} e^{ - ikx}.
\end{eqnarray}
This second   Fourier integral is  calculated  with Filon,
considering again
 $N=20001$ equidistant mesh points  in the interval $[-k_{max},k_{max}]$.

In order to obtain in general
 a quantitative estimate for arising numerical uncertainties,
 we define in this context the quantity
\begin{equation}\label{delta2}
\epsilon = \frac{||R^{num}(x)-R(x)||^2}{||R^{num}(x)||^2}
\end{equation}


with $R^{num}$ denoting  our numerical result for the real part of
the
 response derived
 via the inversion formula (\ref{inversion}). Concerning the norm $||$, we use
  (\ref{int6}), i.e.\ the $L_2$ norm.

\begin{center}
\begin{figure}[btp]
\vspace{0.5cm}
\centerline{\psfig{figure=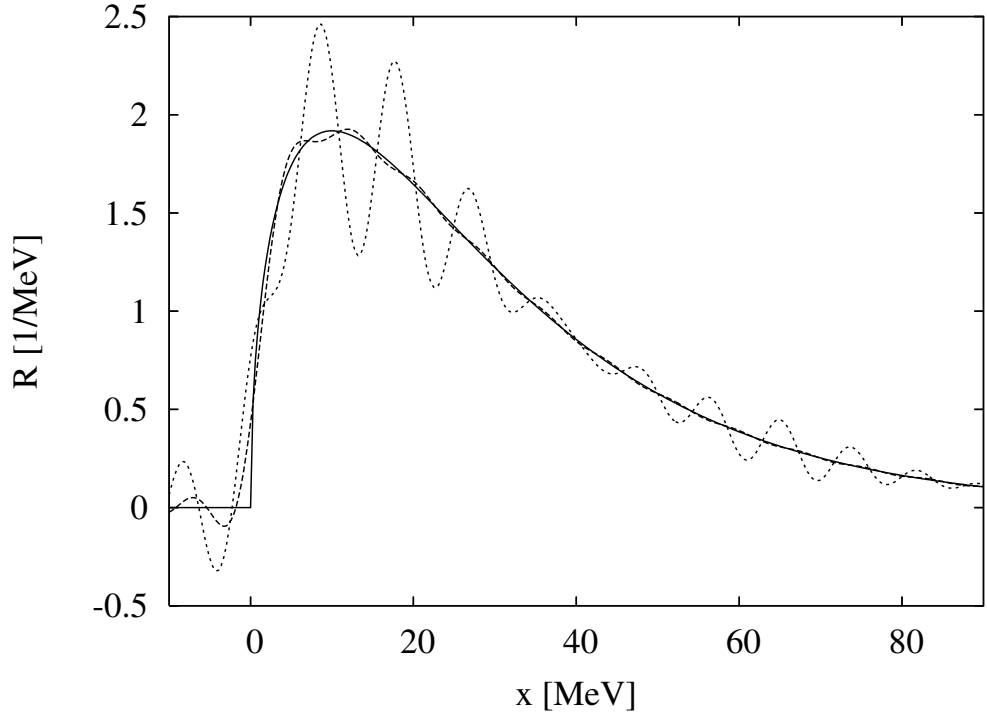,width=14cm,angle=270}}
\vspace{0.5cm} \caption{Full curve: Exact Response function
(\ref{lor5}), dashed curve:
 Response function yielded by inversion formula (\ref{inversion}) (calculation
 A) with  $k_{max}=0.8$ MeV$^{-1}$. Dotted curve: result for response
 in calculation B with  $k_{max}=0.8$ MeV$^{-1}$.
  In all curves, $\sigma_i$ = 10 MeV has been used.}
\label{fig1}
\end{figure}
\end{center}

In our first attempt, calculation {\bf A}, we choose $k_{max} \equiv
  k^{A}_{max}=0.8$
 $\mbox{MeV}^{-1}$
 in
 (\ref{inverse2}).  The resulting response ($\epsilon=2.2 \,\,  10^{-3}$, dashed
 curve in Fig.\ \ref{fig1}),
  using (\ref{inverse2}),  turns out to be qualitatively
   correct beyond the threshold region,
 but unsatisfactory at threshold. The reason for this failure and a
  convenient improvement will be discussed below. Apart from that point,
 we can  conclude that our new inversion formula (\ref{inversion})
  works, at least in principle for a LIT with arbitrarily high precision.
In practical applications, however,  the LIT  $L(y)$ is not known
  perfectly well.
  In order to take this  important fact into account in the present case study,
  we proceed as
 follows.  We take, for a given value of $y$, the exact Lorentz transform
 of (\ref{lor5})
 (``exact'' within the limits of the compiler precision)  in
 the form $0.x_1x_2x_3x_4x_5...\,\,\,10^{-d}$ with $x_i,d$ integer numbers
and $x_1 \neq 0$.
 Then, we substitute this number by hand according to
\begin{equation}\label{mod}
0.x_1x_2x_3x_4x_5x_6...\,\,\, 10^{-d} \quad \longrightarrow
\quad 0.x_1x_2x_3\,\,\, 10^{-d} \,\, ,
\end{equation}
i.e. we cut all relevant digits beyond the first three ones.
This ad hoc procedure simulates a numerical error in $L$ in the range
 $\sim 10^{-3} - 10^{-2}$    -- a level which can, under favourable conditions,
  reached in
 state-of-the-art calculations in nuclear physics.  This modified Lorentz
 transformation, used from now on in  this work,
   can hardly be distinguished from the original one.
 However,
 the resulting response function, yielded by the application of
 (\ref{inversion}), shows large  oscillations
 ($\epsilon=5\, \, 10^{-2}$,
 dotted curve in Fig.\ \ref{fig1}). This calculation is denoted from now
 on as  {\bf calculation B}.
 This numerical fact shows insistently the well-known
 ill-posed character of the Lorentz
 integral transformation: a small change in $L$ (lhs. of (\ref{int1})), generated
 in our test case by the modification (\ref{mod}),  leads
 to large changes in the response $R$ (rhs. of equation (\ref{int1})).

\begin{center}
\begin{figure}[btp]
\vspace{0.5cm}
\centerline{\psfig{figure=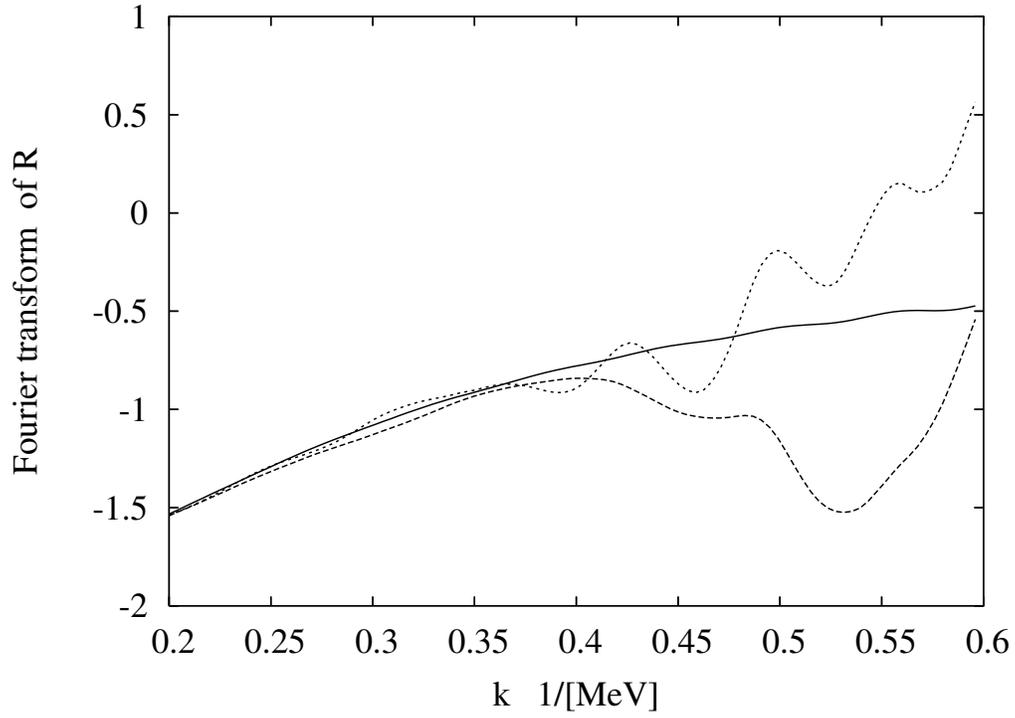,width=14cm,angle=270}}
\vspace{0.5cm} \caption{
 Real part of ${\tilde
R}(k)$ in calculation B for
 $\sigma_i =10$ MeV  (dashed) and $\sigma_i=11$ MeV (dotted).
 The full curve corresponds to the result for calculation A with
 $\sigma_i = 10$ MeV.}
\label{fig2}
\end{figure}
\end{center}

As a novel feature,
 our inversion procedure allows now to pin down precisely
 the source for the ill-posed
  character of the Lorentz transform. For that purpose,
 Fig.\ \ref{fig2}  shows for calculation A and B the
 real part of the  corresponding Fourier
 transforms ${\tilde R}(k)$
 yielded via (\ref{int3}), i.e.\
\begin{equation}\label{int12}
 {\tilde R}(k)=
\frac{1}{\sqrt{2\pi}} \frac{{\tilde L}(k)}{{\cal K}_{LIT}(k)}\, .
\end{equation}
 One obtains a remarkable fact. For  arguments
 $k>0.4$ MeV$^{-1}$,
 the Fourier transform yielded in calculation B (dashed) differs
 considerably  from  the ``exact'' result A (full).
 This can be easily understood
 by the explicit form of ${\tilde K}_{LIT}(k)$ (\ref{lor2}).
 Its inverse is growing
 exponentially  so that in (\ref{int12})  even small errors in ${\tilde L}(k)$
are tremendously amplified. This fact also explains intuitively why for small
 arguments $k$ both Fourier transforms are almost identical. This
 important result
 can also be understood from a more rigorous point of view. For $k \rightarrow
 0$, we have
\begin{equation}\label{eq101}
\lim_{k \rightarrow 0} \sqrt{2 \pi} {\tilde R}(k) = \int dx R(x) = \int dy L(y)
 \,\, ,
\end{equation}
 where the last equation follows from the structure of Lorentz kernel
(\ref{LIT}). In consequence, if the LIT $L$ contains only a small error, then
 the last integral in (\ref{eq101}) will also contain only a small error.
 Therefore,  the Fourier transform ${\tilde R}(k)$ can be determined
 with great precision  for $k \rightarrow 0$ via (\ref{int3}) -- despite
 that the LIT is ill-posed.

Let us summarize this important result again: The LIT is an ill-posed
 integral transform. However, this ill-posed character does not affect
 at all the numerical stability of the obtained  Fourier transform
 ${\tilde R}(k)$  of the resulting response for sufficiently
 small arguments $k$.  Only for
 moderate and large arguments $k$, numerical instabilities arise which can
 be traced back to  the ill-posed character of the LIT.

 {\it In consequence:  If it were possible
 to determine, from general principles, the asymptotic behaviour of the
 Fourier transform ${\tilde R}(k)$, one could hope to circumvent, or at least
 to reduce significantly, any  arising numerical instability
 problems due to the ill-posed character of the LIT.} From a certain point of
 view, the ill-posed character of the LIT could be ``healed''.

For our case study, we        proceed now as follows. At first, we have
 to  fix the value $k_{thres}$ up to which the ill-posed character
 is irrelevant for the determination of ${\tilde R}(k)$  via (\ref{int12}).
 Unlike as in our case study, the exact Fourier transform   ${\tilde R}(k)$
 is of course not known in general. In practice, one could help oneself
 by repeating the calculation for a slightly different
 value $\sigma_i$.  In an ideal calculation, without any
numerical errors,  the resulting Fourier transform ${\tilde R}(k)$ should of course
 be independent from the chosen value $\sigma_i$.
 In Fig.\ \ref{fig2}  the real part of   ${\tilde R}(k)$ can
 be compared for our standard value $\sigma_i =10$ MeV (dashed)
  to the one for   $\sigma_i =11$ MeV (dotted).
 Till about $k=0.25$ MeV$^{-1}$,
  both Fourier transforms are almost identical, however, 
   especially for
 arguments  larger than about $k=0.4$ MeV$^{-1}$
 substantial differences arise.
  Therefore, we could trust our results only
 till about  $k_{thres}=0.25$  MeV$^{-1}$.  For
$k > 0.25$ MeV$^{-1}$
 we have to substitute the unphysical result obtained in calculation
 B by a more reasonable one.
For that purpose, we exploit
  general theorems on the
 asymptotic behaviour of Fourier transforms \cite{Lig58,BrS70}. It turns out
 that quite in general the latter is unambigiously fixed by the knowledge
 of the threshold behaviour of the response function $R$.
 Now, the threshold behaviour of $R$ is, for a given reaction, in general 
 known from
  basic  physical principles.  Without any considerable loss of generality,
  we can therefore assume that we know  not only in our case study, but
 quite  in general
  the threshold behaviour of our unknown  response (\ref{lor5}), i.e.\
\begin{equation}\label{thres1}
R(x) \sim \sqrt{x} \,  \, .
\end{equation}
 Due to \cite{BrS70}, its Fourier transform must therefore  behave
 asymptotically  as
\begin{equation}\label{thres2}
\tilde R(k) \sim \frac{1}{\sqrt{k}^3} \,  \, ,
\end{equation}
   in agreement  with
 the exact analytical result (\ref{fourier_ana1}).

\begin{center}
\begin{figure}[btp]
\vspace{0.5cm}
\centerline{\psfig{figure=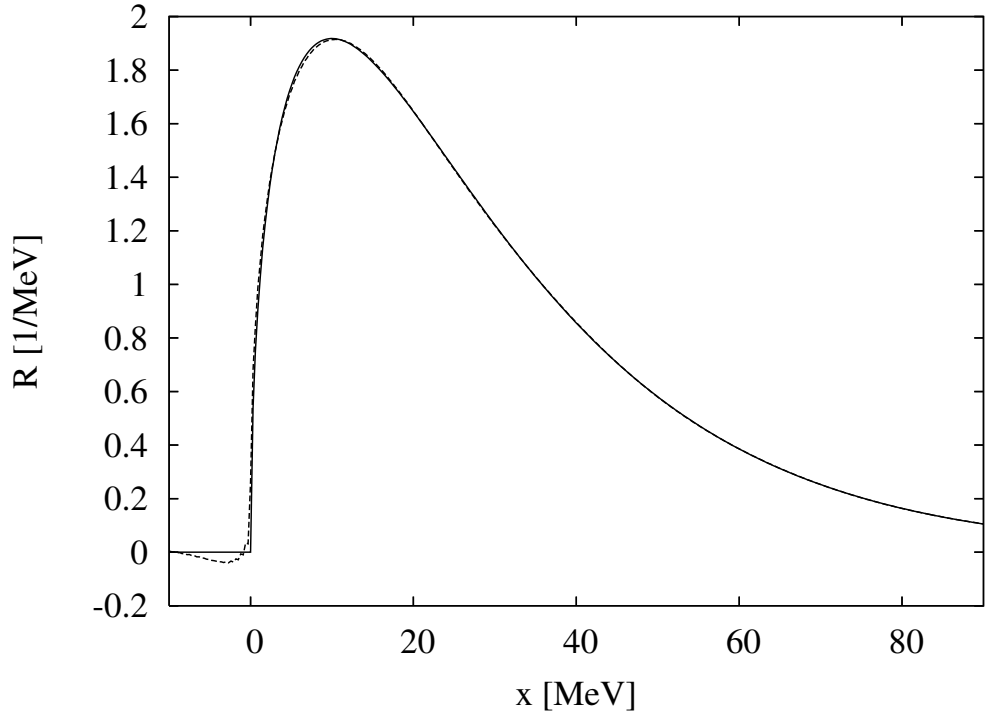,width=14cm,angle=270}}
\vspace{0.5cm} \caption{Full curve: exact response function
(\ref{lor5}), dashed curve:
 response yielded in calculation C with $\sigma_i$=10 MeV.}

\label{fig3}
\end{figure}
\end{center}

Now, it is almost straightforward how to proceed. Only for $k < k_{thres}$
 we use in (\ref{inverse2}) numerical results for $\tilde L(k)$. For
 $k > k_{thres}$, we use  instead the asymptotic ansatz (compare with
(\ref{int3}))
\begin{equation}\label{asy4}
{\tilde L}_{asy}(k) =   \sqrt{2 \pi} \tilde K(k) \tilde R_{asy}(k)
\end{equation}

 with
\begin{equation}\label{asy5}
\tilde R_{asy}(k) = \frac{c}{\sqrt{k}^3} \,\, . 
\end{equation}
The complex parameter $c$ is choosen to guarantee a continous integrand at
 $k= \pm k_{thres}$ in (\ref{inverse2}). By this prescription, it is of course
 easily possible to extend the limit $k_{max}$ in (\ref{inverse2}) to
 arbitrarily large values without any arising numerical errors.
 In the resulting calculation {\bf C}, we choose $k_{max}=10$ MeV$^{-1}$
  with again
 $N=20001$ equidistant mesh points in the region $[-k_{max},k_{max}]$.
 The resulting response is depicted as  the dashed  curve in Fig.\
 \ref{fig3} ($\epsilon = 4.2 \,\,  10^{-4}$).
 One readily realizes two very important improvements:
 (i) The heavy oscillations found in calculation B have vanished, (ii)
 the  threshold behaviour  is considerably improved compared to
 calculation A. Both features can be traced back to the improved treatment of
 $\tilde L(k)$  by exploiting the known asymptotic
 behaviour of $\tilde R(k)$.   Since we assume that we know the threshold
 behaviour of the response,
 the still arising problem of a small, but
nonvanishing numerical result for $R(x)$ below threshold is irrelevant
 in practice.

 In general, for an unknown response,
    it is of course not clear whether the asymptotic form is already working
 for $k = k_{thres}$ where the Fourier transforms $\tilde R(k)$ start to
 diverge from each other for different choices of $\sigma_i$.
  The present case, considers, from that point of view,
 even an ``almost'' worst case scenario,
since the Fourier transform is decreasing
  rather slowly, ${\tilde R}(k) \sim \frac{1}{\sqrt{k}^3}$.
  It turns out that
  for $k =0.8$ MeV$^{-1}$, the 
 asymptotic form and the exact form
 of the real part of the Fourier transform, (\ref{asy5}) vs.
 (\ref{fourier_ana1}), differ by not less than about
 40 percent. Nevertheless, even this quite reduced precision in the
 knowledge of $\hat R(k)$ for $k$ at quite moderate values beyond 
 $k_{thres}$
 is absolutely
 sufficient to obtain reasonable  results in the inversion of the LIT.
   If the Fourier transform  decreases
 stronger, e.g.\ exponentially, one can expect even more
 reliable results within our proposed inversion technique.

 \section{Summary}

   We illuminated the ill-posed character of the LIT approach by
   providing an exact inversion formula and giving the spectrum
  of the LIT kernel. The exact inversion formula requires the Fourier transform of the LIT transformed response, which has to be obtained numerically,
  whereas the Fourier transform  of the LIT kernel is analytically
  known. Since $ \tilde K^{-1}( k) $ occurs in the inversion formula,
      even small numerical errors in $ \tilde L(k)$ at large
      $k$-values are drastically enhanced which lead to violent oscillations of
      $R(x)$ in the inverse Fourier transformation. We illustrated
      that situation in a numerical test case. Evaluating  $L(y)$
      and its FT very precisely and using the inversion  formula
      we recovered the model response $R(x)$ with satisfactory
      precision, at least  beyond the threshold region.
  However, truncating $L(y)$ to about three
      digits revealed very drastically the ill-posed character of
      LIT, namely leading to wild oscillations in $R(x)$
      evaluated by the inversion  formula. It is not  the
      behaviour  of $ \tilde L(k)$ for small $k$-values which
      causes the oscillations but the behaviour  for the large
      $k$-values. This  is clearly demonstrated by putting $ \tilde
      L(k)$ to its known asymptotic behaviour for $k$ larger a certain
      $k_{thres}$.   The resulting $ R(x)$
      gained through the inversion  formula is close to the
      exact one, without oscillations.  In this framework, also
 the threshold behaviour of $R(x)$ can be reproduced satisfactorily.

As not oulined in this work, 
 we verified in addition that for a more structured model response with
 two peaks, like the one used in \cite{AnL05}, 
 that inversion formula works equally well, provided
 one corrects again the asymptotic behavior of $\tilde L(k)$.

        In the applications of LIT to electro weak processes in
        nuclear  physics, $ L(y)$  can be determined in various
 manners, see \cite{EfL07} for further details.
          Using  our inversion
        formula,  the Fourier transformed  $ \tilde L(k)$ of a numerically
   gained LIT is required.
  We conjecture that our results  lead to the  requirement
        to determine $ L(y)$  with such a high
        precision   that its Fourier transform is well under control
          for small and intermediate arguments, but not necessarily for
 large arguments.
        If this is guaranteed, and if the threshold behaviour of the 
     response is known,   the proposed
 method allows  to heal the ill-posed character of the LIT {\it without any
 use of regularization techniques}. From a conceptual point of view,
this a very remarkably result.

 Since our  proposed ansatz is quite general,
 provided the integral kernel is given in the form (\ref{kernel}),  one can
 apply the techniques developed in this work to a variety of different
 other ill-posed problems  in physics and engineering.

\end{document}